\documentclass[twocolumn,aps,pra,showpacs]{revtex4}
\usepackage[T1]{fontenc}
\usepackage[latin9]{inputenc}
\usepackage{amsmath}
\usepackage{amssymb}
\usepackage{graphicx}
\usepackage{esint}

\makeatletter
\@ifundefined{textcolor}{}
{%
 \definecolor{BLACK}{gray}{0}
 \definecolor{WHITE}{gray}{1}
 \definecolor{RED}{rgb}{1,0,0}
 \definecolor{GREEN}{rgb}{0,1,0}
 \definecolor{BLUE}{rgb}{0,0,1}
 \definecolor{CYAN}{cmyk}{1,0,0,0}
 \definecolor{MAGENTA}{cmyk}{0,1,0,0}
 \definecolor{YELLOW}{cmyk}{0,0,1,0}
}

\makeatother

\begin{document}

\title{Radio-frequency spectroscopy of a strongly interacting spin-orbit
coupled Fermi gas}

\author{Zhengkun Fu$^{1}$, Lianghui Huang$^{1}$, Zengming Meng$^{1}$,
Pengjun Wang$^{1}$, Xia-Ji Liu$^{2}$, Han Pu$^{3,4}$, Hui Hu$^{2}$,
and Jing Zhang$^{1}$$^{\dagger}$}

\affiliation{$^{1}$State Key Laboratory of Quantum Optics and Quantum Optics
Devices, Institute of Opto-Electronics, Shanxi University, Taiyuan
030006, P. R. China \\
 $^{2}$Centre for Atom Optics and Ultrafast Spectroscopy, Swinburne
University of Technology, Melbourne 3122, Australia\\
 $^{3}$Department of Physics and Astronomy, and Rice Quantum Institute,
Rice University, Houston, TX 77251, USA\\
$^4$Center for Cold Atom Physics, Chinese Academy of Sciences, Wuhan 430071, China}

\date{\today}
\begin{abstract}
We investigate experimentally and theoretically radio-frequency spectroscopy
and pairing of a spin-orbit-coupled Fermi gas of $^{40}$K atoms near
a Feshbach resonance at $B_{0}=202.2$ G. Experimentally, the integrated
spectroscopy is measured, showing characteristic blue and red shifts
in the atomic and molecular responses, respectively, with increasing
spin-orbit coupling. Theoretically, a smooth transition from atomic
to molecular responses in the momentum-resolved spectroscopy is predicted,
with a clear signature of anisotropic pairing at and below resonance.
Our many-body prediction agrees qualitatively well with the observed
spectroscopy near the Feshbach resonance.
\end{abstract}

\pacs{05.30.Fk, 03.75.Hh, 03.75.Ss, 67.85.-d}

\maketitle

\section{Introduction}

Owing to the unprecedented controllability of interaction and dimensionality,
strongly interacting ultracold Fermi gases have proven to be an ideal
desktop system in the study of pairing and superfluidity \cite{Bloch2008,Giorgini2008}.
Using magnetic field Feshbach resonances, a crossover from Bose-Einstein
condensates (BECs) to Bardeen-Cooper-Schrieffer (BCS) superfluids
was successfully demonstrated in 2004 \cite{Regal2004,Zwierlein2004}
and the pairing properties at the crossover have been characterized
in a number of means since then, including particularly radio-frequency
(rf) spectroscopy \cite{Chin2004,Schunck2008,Stewart2008}. The latest
development in this field is the realization of a synthetic spin-orbit
coupling, which couples the pseudo-spin of neutral atoms to their
orbital motion \cite{Lin2011,Wang2011-PRA,Chen2012,exptShanXi,exptMIT}.
Such a spin-orbit coupling is responsible for a variety of intriguing
phenomena in different fields of physics. A well-known example is
the recently discovered topological insulators in solid-state \cite{Qi2010,Hasan2010}.
In the context of ultracold atomic Fermi gases, it is therefore natural
to ask: what is the consequence of the interplay of strong interaction
and spin-orbit coupling?

In this paper, we investigate rf-spectroscopy of a strongly interacting
spin-orbit coupled Fermi gas of $^{40}$K atoms. The previous works
on spin-orbit coupled Fermi gas explored essentially the non-interacting
limit \cite{exptShanXi,exptMIT}. The current work is the first demonstration
of effects of spin-orbit coupling in an {\em interacting} Fermi gas. In recent BEC-BCS
experiments, rf-spectroscopy has been particularly useful in studying
pairing and superfluidity, yielding information about the pairing gap \cite{Chin2004}
and pair size \cite{Schunck2008}. Furthermore, momentum-resolved
rf-spectroscopy gives a direct information of the low-energy excitation
spectrum and quasiparticles \cite{Stewart2008}. Here, by developing
a many-body \textit{T}-matrix theory we show theoretically that both
atomic and molecular responses in the rf-spectroscopy, arising respectively
from free fermionic atoms and bosonic molecules, are greatly modified
by spin-orbit coupling. In particular, the resulting anisotropic pairing,
dominated by the two-body effect on the BEC side of the Feshbach resonance and
by the many-body effect near resonance, is clearly evident in the
momentum-resolved spectroscopy. Experimentally, we measure the integrated
rf-spectroscopy and report characteristic blue and red shifts in the
atomic and molecular responses, respectively, which are in good agreement with theory.

The remainder of this paper is organized as follows. In Sec. II, we
describe briefly the experimental setup and the model Hamiltonian.
In Sec. III, we present the experimental and theoretical results of
radio-frequency spectroscopy near Feshbach resonances. We introduce
briefly a many-body \textit{T}-matrix theory in Sec. III(A), and in
Sec. III(B) and Sec. III(C) we discuss respectively the integrated
spectroscopy of bound molecules and the momentum-resolved spectroscopy
of fermionic pairs at resonance. The comparison between experiment
and theory is reported in Sec. III(D). Finally, we conclude in Sec.
IV. The Appendix A is devoted to solving the many-body \textit{T}-matrix
theory within the pseudogap approximation.

\section{Experimental setup and model Hamiltonian}

The experimental setup has been described in our previous works \cite{exptShanXi,Xiong2008,Xiong2010,Wang2011},
in which a Bose-Fermi mixture of $^{40}$K and $^{87}$Rb atoms is cooled using well-developed evaporative
and sympathetic cooling techniques in a quadrupole-Ioffe configuration
magnetic trap, and is transported into an optical trap. A degenerate
Fermi gas of about $N\simeq2\times10^{6}$ $^{40}$K atoms in the
$|F=9/2,m_{F}=9/2\rangle$ internal state is evaporatively cooled
to temperature $T/T_{F}\simeq0.3$ with bosonic $^{87}$Rb atoms,
where $T_{F}$ is the Fermi temperature defined by $T_{F}=E_{F}/k_{B}=(6N)^{1/3}\hbar\bar{\omega}/k_{B}$
and $\overline{\omega}\simeq2\pi\times130$ Hz is the geometric mean
trapping frequency. $^{87}$Rb atoms in the mixture are then removed
by a 780 nm laser pulse. Subsequently, fermionic atoms are transferred
into the lowest hyperfine state $|F=9/2,m_{F}=-9/2\rangle$ via a
multi-photon rapid adiabatic passage induced by a radio frequency
field at lower magnetic field. To prepare a two-component $^{40}$K
Fermi gas in an equal mixture of $\left|\uparrow\right\rangle =|F=9/2,m_{F}=-7/2\rangle$
and $\left|\downarrow\right\rangle =|F=9/2,m_{F}=-9/2\rangle$ states,
a homogeneous bias magnetic field, produced by the quadrupole coils
(operating in the Helmholtz configuration), is raised to about $B\approx219.4$
G and then a radio frequency ramp around 47.45 MHz is applied for
50 ms. To create strong interactions, the bias field is ramped from
204 G to a value near the $B_{0}=202.2$ G Feshbach resonance at a
rate of about 0.08 G/ms.

\begin{figure}
\begin{centering}
\includegraphics[clip,width=0.48\textwidth]{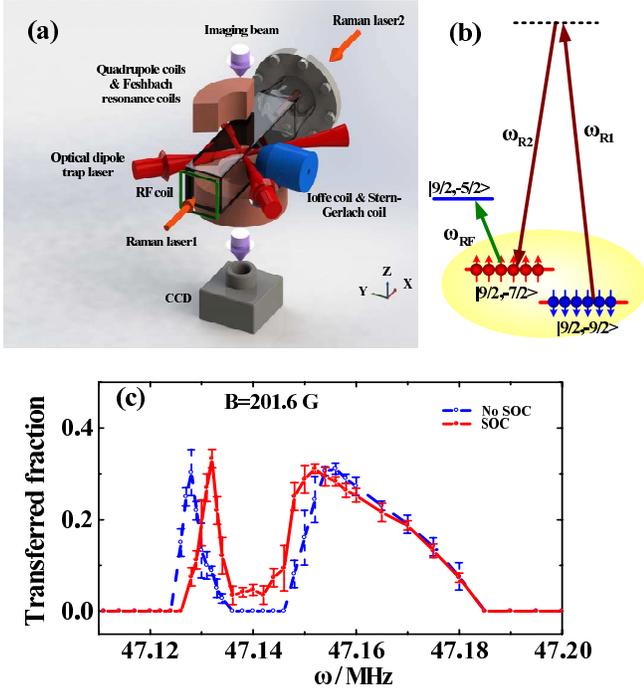}
\par\end{centering}

\caption{(color online) (a) and (b) Experimental realization of a strongly-interacting
Fermi gas of $^{40}$K atoms with spin-orbit coupling. (c) The integrated
rf-spectroscopy below the Feshbach resonance (at $B=201.6$ G and
$a_{s}\simeq2215.6a_{B}$, where $a_{B}$ is the Bohr radius), in
the presence (solid circles) and absence (empty circles) of the spin-orbit
coupling. The Raman detuning is $\delta=0$. The dimensionless interaction
parameter $1/(k_{F}a_{s})\simeq0.66$. The fraction is defined as
$N_{-5/2}/(N_{-5/2}+N_{-7/2})$, where $N_{-5/2}$ and $N_{-7/2}$
are obtained from the TOF absorption image. SOC: spin-orbit coupling.}

\label{fig1}
\end{figure}

We create spin-orbit coupling using Raman process \cite{Lin2011,exptShanXi},
as illustrated in Figs. 1(a) and 1(b). A pair of Raman beams from
a Ti-sapphire laser counter-propagate along the $\hat{x}$ axis and
couple the two spin states. The intensity of beams is $I=50$ mW and
their frequencies are shifted respectively by 75 and 120 MHz, using
two single-pass acousto-optic modulators. These two Raman beams intersect
in the atomic cloud with $1/e^{2}$ radii of 200 $\mu$m and are linearly
polarized along $\hat{z}$ and $\hat{y}$ axis directions, respectively.
The momentum transferred to atoms during the Raman process is $|\textbf{q}_{R}|=2k_{R}\sin(\theta/2)$,
where $k_{R}=2\pi/\lambda_{R}$ is the single-photon recoil momentum,
$\lambda_{R}=772.1$ nm is the wavelength, and $\theta=180^{o}$ is
the intersecting angle of two Raman beams. In the second quantization,
this Raman process may be described by the term
\begin{equation}
H_{R}=\frac{\Omega_{R}}{2}\int d\mathbf{r}\left[\psi_{\uparrow}^{\dagger}\left({\bf r}\right)e^{i2k_{R}x}\psi_{\downarrow}\left({\bf r}\right)+\textrm{\ensuremath{\mathrm{H.c.}}}\right],
\end{equation}
where $\psi_{\sigma}^{\dagger}\left({\bf r}\right)$ is the creation
field operator for atoms in the spin-state $\sigma=\uparrow,\downarrow$
and $\Omega_{R}$ is the coupling strength of Raman beams. For a detailed
discussion on the Raman coupling strength $\Omega_{R}$, we refer
to the recent theoretical work by Wei and Mueller \cite{Wei2013}.

In this paper, we use a larger bias magnetic field than the one used
in our previous study \cite{exptShanXi}, in order to create strong
interactions. Due to a decoupling of the nuclear and electronic spins,
the Raman coupling strength decreases with increasing the bias field
\cite{Wei2013}. To compensate this reduction, here we use a smaller
detuning of the Raman beams with respect to the atomic ``D1'' transition.

To see clearly the spin-orbit coupling in our setup, it is convenient
to take a gauge transformation, $\psi_{\uparrow}({\bf r)}=e^{ik_{R}x}\Psi_{\uparrow}({\bf r)}$
and $\psi_{\downarrow}({\bf r)}=e^{-ik_{R}x}\Psi_{\downarrow}({\bf r)}$.
Our system may therefore be described by a model Hamiltonian ${\cal H}={\cal H}_{0}+{\cal H}_{int}$,
where
\begin{eqnarray}
{\cal H}_{0} & = & \sum_{\sigma}\int d{\bf r}\Psi_{\sigma}^{\dagger}({\bf r)}\frac{\hbar^{2}\left({\bf \hat{k}}\pm k_{R}{\bf e}_{x}\right)^{2}}{2m}\Psi_{\sigma}({\bf r)}+\nonumber \\
 &  & \frac{\Omega_{R}}{2}\int d\mathbf{r}\left[\Psi_{\uparrow}^{\dagger}\left({\bf r}\right)\Psi_{\downarrow}\left({\bf r}\right)+\textrm{\ensuremath{\mathrm{H.c.}}}\right]\label{eq:hami0}
\end{eqnarray}
is the single-particle Hamiltonian and
\begin{equation}
{\cal H}_{int}=U_{0}\int d{\bf r}\Psi_{\uparrow}^{\dagger}\left({\bf r}\right)\Psi_{\downarrow}^{\dagger}\left({\bf r}\right)\Psi_{\downarrow}\left({\bf r}\right)\Psi_{\uparrow}\left({\bf r}\right)
\end{equation}
describes the contact interaction. In the first line of Eq. (\ref{eq:hami0})
we have used ${\bf \hat{k}}=i\mathbf{\nabla}$, ``$+$'' for $\sigma=\uparrow$
and ``$-$'' for $\sigma=\downarrow$. Using the Pauli matrices
$\sigma_{x}$, $\sigma_{y}$ and $\sigma_{z}$, the single-particle
Hamiltonian may be rewritten in a compact form,
\begin{equation}
{\cal H}_{0}=\int d{\bf r}\Phi^{\dagger}\left[\frac{\hbar^{2}\left(k_{R}^{2}+\hat{\mathbf{k}}^{2}\right)}{2m}+h\sigma_{x}+\lambda k_{x}\sigma_{z}\right]\Phi,
\end{equation}
where the spinor field operator $\Phi({\bf r})\equiv[\Psi_{\uparrow}\left({\bf r}\right),\Psi_{\downarrow}\left({\bf r}\right)]^{T}$.
We have defined a spin-orbit coupling constant $\lambda\equiv\hbar^{2}k_{R}/m$
and an ``effective'' Zeeman field $h\equiv\Omega_{R}/2$.

To create a strongly interacting Fermi gas with spin-orbit coupling,
after the bias magnetic field is tuned to a final value $B$ (which
is varied), we ramp up adiabatically the Raman coupling strength in
$15$ ms from zero to its final value $\Omega=1.5E_{R}$ with Raman
detuning $\delta=0$, where the recoil energy
$E_{R}\equiv\hbar^{2}k_{R}^{2}/(2m)\simeq h\times8.36$ kHz. The
temperature of the Fermi cloud after switching on the Raman beams is
at about $0.6T_{F}$, as in our previous measurement for a
non-interacting spin-orbit coupled Fermi gas \cite{exptShanXi}. The
Fermi energy are $E_{F}\simeq2.5E_{R}$ and the corresponding Fermi
wavevector is $k_{F}\simeq1.6k_{R}$.

\section{Radio frequency spectroscopy}

To characterize the strongly interacting spin-orbit coupled Fermi
system, we apply a Gaussian shape pulse of rf field with a duration
time about 400 $\mu$s and frequency $\omega$ to transfer the spin-up
fermions to an un-occupied third hyperfine state $\left|3\right\rangle =\left|F=9/2,m_{F}=-5/2\right\rangle $.
The Gaussian shape pulse is generated by the voltage-controlled rf
attenuators. The Gaussian envelope hence results in the elimination
of the side lobes in rf spectra. The Hamiltonian for rf-coupling may
be written as,
\begin{equation}
{\cal V}_{rf}=V_{0}\int d{\bf r}\left[e^{-ik_{R}x}\psi_{3}^{\dagger}\left({\bf r}\right)\Psi_{\uparrow}\left({\bf r}\right)+\textrm{H.c.}\right],
\end{equation}
where $\psi_{3}^{\dagger}\left({\bf r}\right)$ is the field operator
which creates an atom in $\left|3\right\rangle $ and $V_{0}$ is
the strength of the rf drive. The effective momentum transfer $k_{R}{\bf e}_{x}$
in ${\cal V}_{rf}$ results from the gauge transformation. After
the rf pulse, we abruptly turn off the optical trap, the magnetic
field and the Raman laser beams, and let the atoms ballistically expand for
12 ms in a magnetic field gradient applied along $\hat{z}$ and take
time-of-flight (TOF) absorption image along $\hat{y}$. We measure
the spin population of the final state $|3\rangle$ for different
rf frequencies to obtain the rf spectra $\Gamma(\omega)$.

For a weak rf drive, the number of transferred fermions can be calculated
using linear response theory. At this point, it is important to note
that the final-state interactions for $^{40}$K atoms in the third
state and in the spin-up or spin-down state is typically small \cite{Bloch2008}.
Theoretically, the rf transfer strength at a given momentum is therefore
determined entirely by the single-particle spectral function of spin-up
atoms ${\cal A}_{\uparrow\uparrow}$:
\begin{eqnarray}
\Gamma({\bf k},\omega) & = & {\cal A}_{\uparrow\uparrow}({\bf k}+k_{R}{\bf e}_{x},\epsilon_{{\bf k}}-\mu-\hbar\omega+\hbar\omega_{3\uparrow})\times\nonumber \\
 &  & f(\epsilon_{{\bf k}}-\mu-\hbar\omega+\hbar\omega_{3\uparrow}),
\end{eqnarray}
where $\epsilon_{{\bf k}}\equiv\hbar^{2}k^{2}/(2m)$, $\mu$ is the
chemical potential of the spin-orbit system, $\hbar\omega_{3\uparrow}\simeq\hbar\times47.1$
MHz is the energy splitting between the third state and the spin-up
state, $f(x)\equiv1/(e^{x/k_{B}T}+1)$ is the Fermi distribution function,
and we have taken the coupling strength $V_{0}=1$. For $^{6}$Li
atoms, however, the final state effect is usually significant \cite{Schunck2008}.
The rf transfer strength will no longer be simply determined by the
single-particle spectral function.

Experimentally, one could measure the momentum-resolved rf
spectroscopy along the \textit{x}-direction
$\Gamma(k_{x},\omega)\equiv\sum_{k_{y},k_{z}}\Gamma({\bf
k},\omega)$, or, after integration obtain the fully integrated rf
spectroscopy $\Gamma(\omega)\equiv\sum_{{\bf k}}\Gamma({\bf
k},\omega)$. Due to small signal to noise ratio, we currently have
difficulty obtaining momentum-resolved rf signal experimentally.

In Fig. 1(c), we show the integrated rf-spectroscopy of an interacting
Fermi gas below the Feshbach resonance, with or without spin-orbit
coupling. Here, we carefully choose the one photon detuning of the
Raman laser to avoid shifting Feshbach resonance by the Raman laser
on the bound-to-bound transition between the ground Feshbach molecular
state and the electronically excited molecular state. The narrow and
broad peaks in the spectroscopy should be interpreted respectively
as the rf-response from free atoms and fermionic pairs. With spin-orbit
coupling, we find a systematic blue shift in the atomic response and
a red shift in the pair response. The latter is an unambiguous indication
that the properties of fermionic pairs are strongly affected by spin-orbit
coupling.

\subsection{Many-body \textit{T}-matrix theory}

Let us now consider theoretical understanding of the observed red
shift for fermionic pairs. Near Feshbach resonances, it is important
to treat atoms and fermionic pairs on an equal footing. For this purpose,
it is convenient to develop a many-body theory within the \textit{T}-matrix
approximation by summing all ladder diagrams \cite{HLDPRA2008,HLDNJP2010}.
In the presence of spin-orbit coupling, it is necessary to define
a finite-temperature Green function
\begin{equation}
{\cal G}({\bf r},{\bf r}^{\prime};\tau>0)\equiv-\left\langle \Phi\left({\bf r,}\tau\right)\Phi^{\dagger}\left({\bf r}^{\prime}{\bf ,}0\right)\right\rangle ,
\end{equation}
which is a $2$ by $2$ matrix even in the normal state. We adopt
a partially self-consistent \textit{T}-matrix scheme and take one
non-interacting and one fully dressed Green function in the ladder
diagrams \cite{HLDNJP2010}. The summation of all ladder diagrams
leads to the Dyson equation,
\begin{equation}
{\cal G}(K)=\left[{\cal G}_{0}^{-1}(K)-\Sigma(K)\right]^{-1},\label{eq:Dyson equation}
\end{equation}
where the self-energy is given by
\begin{equation}
\Sigma(K)=\sum_{Q}\left[t(Q)(i\sigma_{y})\tilde{{\cal G}_{0}}(K-Q)(i\sigma_{y})\right].\label{eq:Self energy}
\end{equation}
Here
\begin{equation}
t(Q)\equiv\frac{U_{0}}{1+U_{0}\chi\left(Q\right)}
\end{equation}
 is the (scalar) \textit{T}-matrix with a two-particle propagator
\begin{equation}
\chi\left(Q\right)=\frac{1}{2}\sum_{K}\text{Tr}\left[{\cal G}(K)\left(i\sigma_{y}\right)\tilde{{\cal G}}_{0}(K-Q)\left(i\sigma_{y}\right)\right]\label{eq:Pair propagator}
\end{equation}
and the non-interacting Green function
\begin{equation}
{\cal G}_{0}(K)=\left[i\omega_{m}-\epsilon_{{\bf k}}+\mu-E_{R}-h\sigma_{x}-\lambda k_{x}\sigma_{z}\right]^{-1}.\label{eq:Free Green function}
\end{equation}
For convenience, we have used the short notations $K\equiv({\bf k},i\omega_{m})$,
$Q\equiv({\bf q},i\nu_{n})$, and $\sum_{K}\equiv k_{B}T\sum_{{\bf k},\omega_{m}}$,
and $\omega_{m}$ and $\nu_{n}$ are respectively the fermionic and
bosonic Matsubara frequencies. We have also defined a Green function
for holes, $\tilde{{\cal G}}(K)\equiv-[{\cal G}(-K)]^{T}$.

Eqs. (\ref{eq:Dyson equation})-(\ref{eq:Free Green function}) generalize
the earlier \textit{T}-matrix diagrammatic theory without spin-orbit
coupling \cite{HLDNJP2010}. Here the modification arising from the
spin-orbit coupling includes the use of Green functions in a general
2 by 2 matrix form and accordingly the appearance of the vertex $i\sigma_{y}$
in Eqs. (\ref{eq:Self energy}) and (\ref{eq:Pair propagator}). The
derivation of these \textit{T}-matrix equations is too technical and
is not the focus of this work. Therefore, we will discuss the details
of the derivation elsewhere.

In general, the self-consistent \textit{T}-matrix equations are numerically
difficult to solve. At a \emph{qualitative} level, however, we may
adopt a pseudogap decomposition for the \textit{T}-matrix \cite{Chen2005}
and obtain a set of coupled equations for the chemical potential and
pairing gap. In the Appendix A, we discuss in detail the pseudogap
approximation. By solving the coupled equations (see, i.e., Eqs. (\ref{eq:Gap equation})
and (\ref{eq:Number equation})), we calculate the single-particle
spectral function
\begin{equation}
{\cal A}_{\uparrow\uparrow}({\bf k},\omega)\equiv-\frac{1}{\pi}\mathop{\rm Im}{\cal G}_{\uparrow\uparrow}(K)
\end{equation}
 and the rf-transition strengths $\Gamma(k_{x},\omega)$ and $\Gamma(\omega)$.
To take into account the experimental energy resolution of the spectroscopy
$\gamma\sim0.1E_{R}$ \cite{exptMIT}, we replace the Dirac delta
function (which appears in ${\cal A}_{\uparrow\uparrow}({\bf k},\omega)$)
by
\begin{equation}
\delta\left(x\right)=\frac{\gamma/\pi}{x^{2}+\gamma^{2}}.
\end{equation}

\begin{figure}
\begin{centering}
\includegraphics[clip,width=0.48\textwidth]{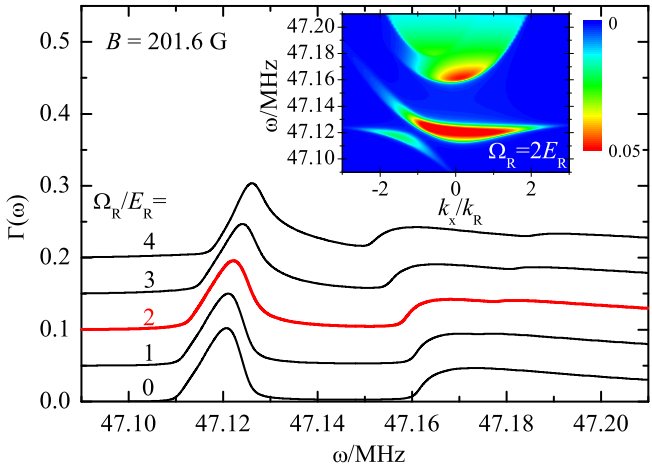}
\par\end{centering}

\caption{(color online) Evolution of the predicted integrated rf-spectroscopy
as a function of the Raman coupling strength. Here, $1/k_{F}a_{s}=0.66$,
$T=0.6T_{F}$, and $k_{F}=1.6k_{R}$. The inset shows a linear contour
plot of the predcited momentum-resolved spectroscopy at $\Omega_{R}=2E_{R}$.}

\label{fig2}
\end{figure}

\subsection{Integrated spectroscopy of bound molecules}

In Fig. 2, we predict the integrated rf-spectroscopy of a spin-orbit
coupled Fermi gas at $B=201.6$ G and $T=0.6T_{F}$. With increasing
the strength of Raman beams from $0$ to $4E_{R}$, the atomic and
pair peaks shift to the higher and lower frequencies, respectively,
in qualitative agreement with the experimental observation in Fig.
1(c). At this interaction parameter ($1/(k_{F}a_{s})\simeq0.66$),
the red shift of pair peaks may be understood from the binding energy
of pairs in the two-body limit: the stronger effective Zeeman field
$h$, the smaller binding energy of two-particle bound states \cite{Jiang2011}.
In the inset, we show the prediction of the momentum-resolved rf-spectroscopy
at $\Omega_{R}=2E_{R}$. It is highly asymmetric as a function of
momentum. The predicted atomic response is in good agreement with
the experimental observation for a non-interacting spin-orbit-coupled
Fermi gas \cite{exptShanXi,exptMIT}.

\begin{figure}
\begin{centering}
\includegraphics[clip,width=0.48\textwidth]{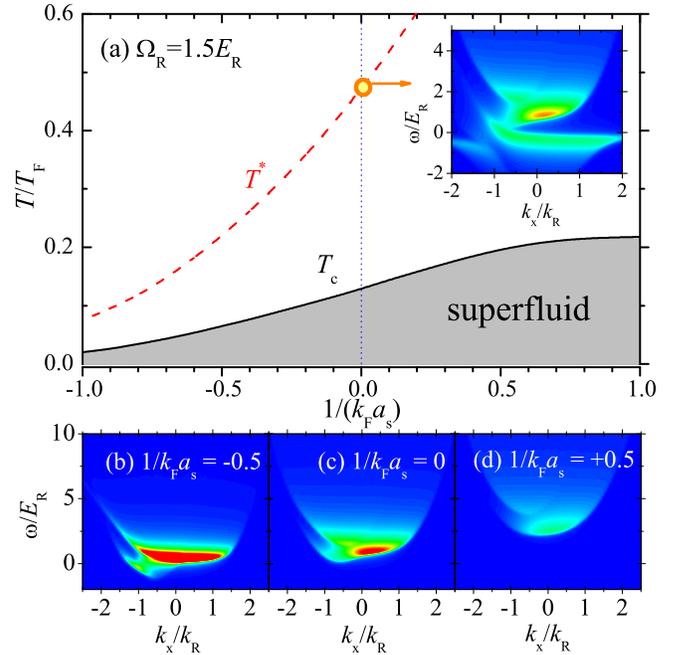}
\par\end{centering}

\caption{(color online) (a) Predicted phase diagram of a strongly-interacting
spin-orbit coupled Fermi gas at $\Omega_{R}=1.5E_{R}$ and $k_{F}=1.6k_{R}$.
(b)-(d) Linear contour plots of the zero temperature momentum-resolved
rf-spectroscopy across the Feshbach resonance, in arbitrary units.
The inset in (a) shows the momentum-resolved rf spectroscopy at the
pseudgap temperature $T^{*}$ in the unitary limit. In theoretical
calculations, we have set $\omega_{3\uparrow}=0$.}

\label{fig3}
\end{figure}

\subsection{Momentum resolved spectrosocpy near Feshbach resonances}

We now turn to the rf-spectroscopy in the vicinity of the Feshbach
resonance. In Fig. 3(a), we plot the superfluid transition temperature
$T_{c}$ and the pairing breaking (pseudogap) temperature $T^{*}$
of a spin-orbit coupled Fermi gas at $\Omega_{R}=1.5E_{R}$ and $k_{F}=1.6k_{R}$.
The pseudogap temperature is calculated using the standard BCS mean-field
theory without taking into account the preformed pairs (i.e., $\Delta_{pg}=0$)
\cite{Chen2005}. We find that the region of superfluid phase is suppressed
by spin-orbit coupling. In particular, at resonance the superfluid
transition temperature is about $T_{c}\simeq0.129T_{F}$, smaller
than the measured value of $T_{c}\simeq0.167(13)T_{F}$ \cite{EoSMIT}
or the predicted value of $T_{c}\simeq0.15T_{F}$ (under the same
pseudogap approximation) for a unitary Fermi gas in the absence of
spin-orbit coupling. Thus, experimentally it becomes more challenge
to observe a spin-orbit coupled fermionic superfluid in the present
experimental scheme.

In Figs. 3(b)-3(d), we show the zero-temperature momentum-resolved
rf-spectroscopy across the resonance. On the BCS side ($1/k_{F}a_{s}=-0.5$),
the spectroscopy is dominated by the response from atoms and shows
a characteristic high-frequency tail at $k_{x}<0$ \cite{exptShanXi,exptMIT,Liu2012}.
Towards the BEC limit ($1/k_{F}a_{s}=0.5$), the spectroscopy may
be understood from the picture of well-defined pairs and shows a clear
two-fold anisotropic distribution \cite{Hu2012}. The spectroscopy
at the resonance is complicated and should be attributed to many-body
fermionic pairs. The change of spectroscopy across resonance is continuous,
in accord with a smooth BEC-BCS crossover \cite{Giorgini2008}.

In the inset of Fig. 3(a), we show the momentum-resolved rf-spectroscopy
at the resonance and at the pseudogap pairing temperature $T^{*}$.
It is interesting that the anisotropic distribution survives well
above the superfluid transition temperature $T_{c}$. An experimental
observation of such a spectroscopy would be a strong indication of
the anisotropic \emph{pseudogap} pairing of a spin-orbit coupled Fermi
gas in its normal state.

\begin{figure*}
\begin{centering}
\includegraphics[clip,width=0.8\textwidth]{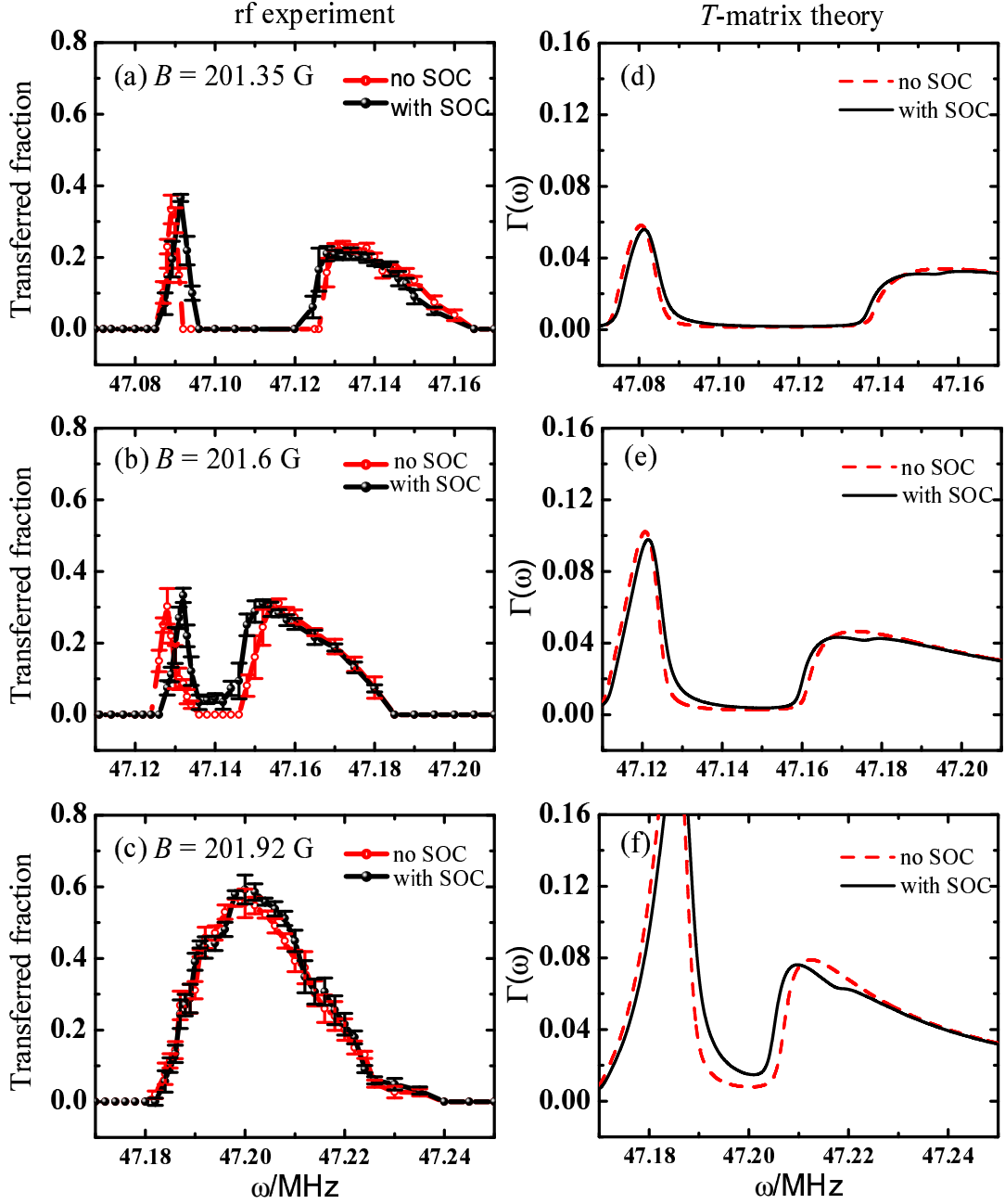}
\par\end{centering}

\caption{(color online) Comparison between theory and experiment for the integrated
rf-spectroscopy. In theoretical simulations, we use $\Omega_{R}=1.5E_{R}$,
$k_{F}=1.6k_{R}$, and $T=0.6T_{F}$, according to the experimental
setup. The solid circles (lines) and empty circles (dashed lines)
show respectively the experimental data (theoretical predictions)
in the presence and absence of spin-orbit coupling with Raman detuning
$\delta=0$. The dimensionless interaction parameter $1/(k_{F}a_{s})$
in (a), (b), and (c) are $0.89$, $0.66$, and $0.32$, respectively. }

\label{fig4}
\end{figure*}

\subsection{Experiment vs Theory}

In Fig. 4, we examine our many-body theory by comparing the theoretical
predictions with the experimental data for integrated rf-spectroscopy.
At the qualitative level, we do not consider the trap effect and take
the relevant experimental parameters at the trap center. Otherwise,
there are no adjustable free parameters used in the theoretical calculations.
As shown in Fig. 4, we find a qualitative agreement between theory
and experiment, both of which show the red shift of the response from
fermionic pairs, as we already discuss in Sec. III(B). The worst agreement
occurs close to the Feshbach resonance (Fig. 4(c)), where our many-body
theory fails to take into account properly the strong interactions between
atoms and pairs.

\section{Conclusion}

In conclusion, we have investigated experimentally and theoretically
rf-spectroscopy and fermionic pairing in a strongly-interacting spin-orbit
coupled Fermi gas of $^{40}$K atoms near a Feshbach resonance. A
red shift of the response from fermionic pairs, induced by spin-orbit
coupling, is observed in integrated rf-spectroscopy below the resonance,
in qualitative agreement with a many-body \textit{T}-matrix calculation.
Momentum-resolved rf-spectroscopy of fermionic pairs has been predicted
across the resonance at all temperatures, showing a characteristic
anisotropic distribution. This is to be confronted in future experiments.

We note that the typical experimental temperature in this work is
about $0.6T_{F}$. In the future, we wish to reduce the temperature
of the strongly interacting Fermi cloud of $^{40}$K atoms down to
$0.2T_{F}$, close to the superfluid transition. This is encourged
by a recent calculation by Wei and Mueller \cite{Wei2013}, who showed
that the heating of the cloud due to Raman transition is not significantly
affected by the magnetic field needed for Feshbach resonances.

Theoretically, it is possible to solve the many-body \textit{T}-matrix
theory without the pseudogap approximation. On the other hand, at
the relatively high temperature $T\sim0.6T_{F}$, alternatively we
may use a virial expansion theory to obtain quantitative predictions
for the radio-frequency spectroscopy in harmonic traps \cite{Liu2009VE,Hu2010VE,Liu2013VE}.
These possibilities will be addressed in later studies.
\begin{acknowledgments}
We thank Zeng-Qiang Yu and Hui Zhai for useful discussions. This research
is supported by NFRP-China (Grant No. 2011CB921601), NSFC Project
for Excellent Research Team (Grant No. 61121064), NSFC (Grant No.
11234008), Doctoral Program Foundation of Ministry of Education China
(Grant No. 20111401130001). XJL and HH are supported by the ARC DP0984637
and DP0984522. HP is supported by the NSF and the Welch Foundation
(Grant No. C-1669).

$^{\dagger}$Corresponding author email: jzhang74@yahoo.com, jzhang74@sxu.edu.cn
\end{acknowledgments}
\appendix
%dummy comment inserted by tex2lyx to ensure that this paragraph is not empty

\section{Pseudogap approximation}

The pseudogap approximation is advanced by the Chicago group \cite{Chen2005}.
In this approximation the \textit{T}-matrix is separated into two
parts, $t(Q)=t_{sc}(Q)+t_{pg}(Q)$, so that the contribution from
the superfluid order parameter for condensed pairs $\Delta_{sc}^{2}$,
\begin{equation}
t_{sc}(Q)=-\frac{\Delta_{sc}^{2}}{T}\delta\left(Q\right),
\end{equation}
and the contribution from the pseudogap for un-condensed pairs,
\begin{equation}
\Delta_{pg}^{2}\equiv-\sum_{Q\neq0}t_{pg}(Q),
\end{equation}
become explicit. The full pairing order parameter is given by $\Delta^{2}=\Delta_{sc}^{2}+\Delta_{pg}^{2}$.
Accordingly, we have the self-energy $\Sigma(K)=\Sigma_{sc}(K)+\Sigma_{pg}(K)$
\cite{Chen2005}, where
\begin{equation}
\Sigma_{sc}=-\Delta_{sc}^{2}(i\sigma_{y})\tilde{{\cal G}_{0}}(K)(i\sigma_{y})
\end{equation}
and
\begin{equation}
\Sigma_{pg}=-\Delta_{pg}^{2}(i\sigma_{y})\tilde{{\cal G}_{0}}(K)(i\sigma_{y}).
\end{equation}
To obtain $\Sigma_{pg}$ in the above equation, it is assumed that
the pair propagator $\chi(Q)$ peaks around $Q=0$ \cite{Chen2005}.

We note that, at zero temperature the pseudogap approximation is simply
the standard mean-field BCS theory, in which
\begin{equation}
\Sigma(K)=-\Delta^{2}(i\sigma_{y})\tilde{{\cal G}_{0}}(K)(i\sigma_{y}).
\end{equation}
Above the superfluid transition, however, it captures the essential
physics of fermionic pairing and therefore should be regarded as an
improved theory beyond mean-field.

To calculate the pseudogap $\Delta_{pg}$, we approximate
\begin{equation}
t_{pg}^{-1}(Q\simeq0)={\cal Z}\left[i\nu_{n}-\Omega_{{\bf q}}+\mu_{pair}\right],
\end{equation}
where the residue ${\cal Z}$ and the effective dispersion of pairs
$\Omega_{{\bf q}}=\hbar^{2}q^{2}/2M^{*}$ are to be determined by
expanding $\chi\left(Q\right)$ about $Q=0$ \cite{Chen2005,footnote}.
The form of $t_{pg}(Q)$ leads to
\begin{equation}
\Delta_{pg}^{2}(T)={\cal Z}^{-1}\sum_{{\bf q}}f_{B}(\Omega_{{\bf q}}-\mu_{pair}),
\end{equation}
where $f_{B}(x)\equiv1/(e^{x/k_{B}T}-1)$ is the bosonic distribution
function. We arrive finally at two coupled equations, the gap equation
\begin{equation}
\frac{1}{U_{0}}+\chi\left(Q=0\right)={\cal Z}\mu_{pair}\label{eq:Gap equation}
\end{equation}
 and the number equation
\begin{equation}
n=\sum_{K}{\cal \textrm{Tr}G}(K),\label{eq:Number equation}
\end{equation}
$ $to determine the superfluid order parameter $\Delta_{sc}$ and
the chemical potential $\mu$, respectively, for a given set of parameters.


\begin{thebibliography}{10}
\bibitem{footnote}At the qualitative level, we approximate the effective
mass of pairs, $M^{*}=2m$.
\end{thebibliography}
\end{document}